\documentclass[11pt]{article}
\pdfoutput=1

\usepackage{amsmath}
\usepackage{amsfonts}
\usepackage{graphicx}

\author{Stefan Janiszewski}
\title{Perturbations of moving membranes in AdS$_7$}

\begin{document}
\maketitle
\begin{center}\emph{Department of Physics, University of Washington, Seattle, WA 98195}\end{center}
\begin{center}stefanjj@u.washington.edu\end{center}
\abstract{We study the stability of uniformly moving membrane-like objects in seven dimensional Anti-de Sitter space. This is approached by a linear perturbation analysis and a search for growing modes. We examine both analytic and numerical configurations previously found in \cite{JK}.}
\newpage

\tableofcontents
\section{Introduction}

The paper \cite{JK} recently explored the properties of defects moving through Anti-de Sitter space. This was a generalization of previous works, \cite{HKKKY,SSG}, which examined the important case of a moving string in AdS$_5$. The focus of \cite{JK} was on the case of a membrane in AdS$_7$ for reasons such as ease in analytics and relevance from a quantum gravity (M-theory) point of view. The major use of such explorations is in the context of the AdS/CFT correspondence \cite{AGMOO}. This duality allows properties of the defects, as calculated in the low energy classical gravity spacetime of AdS, to be translated into information about dual defects probing properties of the conformal field theory. This has led to the determination of quantities like the energy loss of a moving defect, and the screening length between two oppositely charged defects. A well known subclass of the correspondence is that a plasma state of the CFT is dual to an asymptotically AdS space with a Schwarzschild black hole. 

The current paper will examine the classical stability of some of the stationary moving solutions found in \cite{JK} and will focus solely on the case of membranes in AdS$_7$. As discussed in more detail there, this is an interesting case of the correspondence for various reasons. The dual six dimensional CFT is poorly understood, not having a Lagrangian description. The underlying microscopic quantum gravity is M-Theory, allowing work in a less well tested case of the duality. The extended nature of the defect lends hope that it might explore different behavior of the CFT plasma than that probed by the zero dimensional endpoints of strings. The stability of these defects is important in this context; turbulent flow of the CFT plasma may manifest as instability of the probe.

The paper is organized as follows: Section \ref{sec:infsolu} recalls the infinitely extended moving membrane of \cite{JK} and its world-volume metric, and examines perturbations around this stationary solution, including the numerical calculation of the quasinormal modes (after a mathematical detour in Appendix \ref{app:contfrac} into multi-dimensional continued fractions); Section \ref{sec:solu} discusses the special analytic massless mode of the membrane corresponding to the Goldstone mode due to breaking of translation invariance; Section \ref{sec:compact} recalls the compact moving membrane of \cite{JK} and its world-volume metric, and examines the quasinormal modes of perturbations around this stationary solution; Lastly, Section \ref{sec:conclu} concludes with a discussion of the results.

\section{Infinitely extended moving membrane}\label{sec:infsolu}
\subsection{Background metric and equations of motion}

We will consider AdS$_{7}$-Schwarzschild spacetime with metric, $G_{\mu \nu}$:
\begin{equation}\label{eq:metric}
ds^2 = \frac{R^2}{u} \left(-\left(1-u^3\right)dt^2+\frac{du^2}{4u\left(1-u^3\right)}+d\vec{y}^2_5\right)\quad.
\end{equation}
This metric is related to the Anti-de Sitter black hole in the more familiar Poincar\'e coordinates via $u=z^2$. It has the horizon at $u=1$, and the boundary of the AdS space with radius of curvature $R$ is at $u=0$.

The equations of motion for the membrane are determined by the determinant of its induced metric, $g_{ab}\equiv G_{\mu\nu}X^\mu_{,a} X^\nu_{,b}$, where $X^\mu(a)$ is the world-volume map into AdS$_7$, and $f(a)_{,b}$ denotes partial differentiation. From the Nambu-Goto-like action, $S=-T_0 \int \prod_i da_i \sqrt{-g}$, where $T_0$ is the membrane's tension, we obtain the equation of motion:
\begin{equation}
\left(\sqrt{-g}g^{ab}G_{\mu \nu} X^\nu_{,b}\right)_{,a} = 0\quad.
\end{equation}

\subsection{Uniformly moving solution}
\begin{figure}%
\centering
\includegraphics[width=.7\columnwidth]{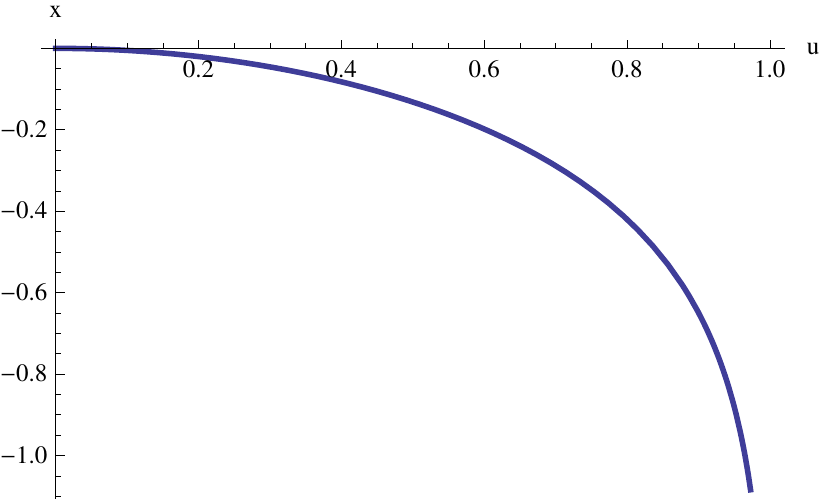}%
\caption{The $x$ profile of the uniformly moving extended sheet as a function of the bulk radius $u$.}%
\label{fig:fig1}%
\end{figure}

For a membrane extended along one of the transverse directions, $y$, and having a non-trivial profile in only one transverse direction, $x$, we can parametrize the world-volume map as $X^\mu(a)=(t,u,y,x(t,u,y),\vec{z}=\vec{z}_0)$ where $\vec{z}$ are the transverse directions upon which the map does not depend. For the ansatz of a stationary solution with no transverse dependence, $x=vt+x(u)$, the equation of motion reduces to:
\begin{equation}\label{eq:eqmo1}
x^2_{,u} = \frac{C^2u^2}{4\left(1-u^3\right)^2}\left(\frac{1-v^2-u^3}{1-(1+C^2)u^3}\right)\quad,
\end{equation}
where $C$ is a constant that is determined by demanding that $x^2_{,u}$ and $-g$ remain positive on the whole world-volume. This requires $C^2=\gamma^2-1$, where $\gamma$ is the Lorentz factor related to the ``velocity'' $v$. This reduces (\ref{eq:eqmo1}) to:
\begin{equation} 
x_{,u} =\pm\frac{v}{2}\frac{u}{1-u^3}\quad,
\end{equation} 
which has the solution:
\begin{equation}
x_v(u)=\pm\frac{v}{6}\left(\log\left(\sqrt{1+u+u^2}\right)-\log(1-u)-\sqrt{3}\arctan\left(\frac{1+2u}{\sqrt{3}}\right)\right)\quad,
\label{eq:statsolu}
\end{equation}
and is shown in Figure \ref{fig:fig1}. The sign is fixed to be the lower one by examining the loss of energy into the black hole as the membrane moves along with the constant velocity $v$.

\subsection{World-volume metric and horizon}
In the original coordinates the world-volume metric $g_{ab}$ is non-diagonal. By reparametrizing:
\[ r \equiv u\gamma^{2/3}, \qquad \tau \equiv \gamma^{-2/3}\left(t+f(u)\right), \qquad w\equiv y \gamma^{1/3}\quad,
\]
where
\[
f'(u) = \frac{g_{tu}}{g_{tt}} = \frac{v\gamma^2 x_{,u}}{(-1+r^3)}\quad,
\]
we have the non-zero components of the metric:
\begin{equation}\label{eq:wvmetric1}
g_{\tau\tau} = \frac{R^2}{r}\left(-1+r^3\right),\qquad g_{rr} = \frac{R^2}{4r^2\left(1-r^3\right)},\qquad g_{ww} = \frac{R^2}{r}\quad.
\end{equation}
In this form it is apparent that there is a horizon on the world-volume at the radial coordinate $u_c=\gamma^{-2/3}$, where $r=1$ and $g_{\tau\tau}$ and $g_{rr}$ change sign.

\subsection{Expansion about stationary solution}

We will now consider perturbations to the above stationary solution. Again, the equations of motion will be determined by an action proportional to the square root of the determinant of the full induced metric, denoted $h_{ab}$. For a world-volume map $ X^\mu(a) = X^\mu_v(a)+\epsilon^\mu(a) $, with $ X^\mu_v$ the stationary solution above, and taking the perturbation $ \epsilon^\mu $ and its derivatives to be small, we can expand the action of the perturbed solution to second order:
\begin{equation}\label{eq:pertaction}
\sqrt{-h} \approx \sqrt{-g}+\epsilon^\mu_{,a}\left.\left(\frac{\partial\sqrt{-h}}{\partial X^\mu_{,a}}\right)\right\vert_{X^\mu_v}+\frac{\epsilon^\mu_{,a}\epsilon^\nu_{,b}}{1+\delta^\mu_\nu \delta^a_b}\left.\left(\frac{\partial}{\partial X^\nu_{,b}}\left(\frac{\partial\sqrt{-h}}{\partial X^\mu_{,a}}\right)\right)\right\vert_{X^\mu_v}\qquad.
\end{equation}

The linear term can be rewritten as:
\[
\epsilon^\mu_{,a}\left.\left(\frac{\partial\sqrt{-h}}{\partial X^\mu_{,a}}\right)\right\vert_{X^\mu_v} = \epsilon^\mu_{,a}\left.\left(\sqrt{-h}h^{ab}G_{\mu \nu} X^\nu_{,b}\right)\right\vert_{X^\mu_v} = \epsilon^\mu_{,a}\left(\sqrt{-g}g^{ab}G_{\mu \nu} X^\nu_{v,b}\right)\qquad,
\]
which vanishes after an integration by parts and use of the equation of motion for $X^\mu_v$. 
 
The quadratic term can be rewritten as:
\[
\frac{\epsilon^\mu_{,a}\epsilon^\nu_{,b}}{1+\delta^\mu_\nu \delta^a_b}\left.\left(\frac{\partial}{\partial X^\nu_{,b}}\left(\frac{\partial\sqrt{-h}}{\partial X^\mu_{,a}}\right)\right)\right\vert_{X^\mu_v} =
\]
\[
\frac{\epsilon^\mu_{,a}\epsilon^\nu_{,b}}{1+\delta^\mu_\nu \delta^a_b}\sqrt{-g}\left(G_{\mu \nu} g^{ab}+G_{\mu \delta}G_{\nu \gamma}X^\delta_{v,c}X^\gamma_{v,d}\left(g^{ac}g^{bd}-g^{ad}g^{bc}-g^{ab}g^{cd}\right)\right)\qquad.
\]

\subsubsection{Transverse perturbations}

For a perturbation only in one direction transverse to the direction of motion we have $\epsilon^\mu = \epsilon \delta^\mu_z$, and the action for the perturbation becomes:
\[
S \sim \frac{\epsilon_{,a}\epsilon_{,b}}{1+\delta^a_b}\sqrt{-g}G_{zz}g^{ab}\qquad.
\]

Parametrizing the stationary world-volume with $(\tau,r,w)$ gives a diagonal $g^{ab}$ and simplifies the covariant equation of motion for the transverse perturbation:
\begin{equation}\label{eq:eqmotrans}
\left(\epsilon_{,b}\sqrt{-g}G_{zz}g^{ab}\right)_{,a}=  0\qquad.
\end{equation}

For transverse perturbations of the form of a plane wave in the $w$ direction, $\epsilon(\tau,r,w) = \epsilon(r)e^{\imath\left(kw-\omega\tau\right)}$, the equation of motion becomes the second order ODE:
\[
\epsilon''(r)+\epsilon'(r)\left(-\frac{1}{r}-\frac{1}{1-r}+\frac{1}{r-\phi}+\frac{1}{r-\phi^2}\right) \] \begin{equation}+\frac{\epsilon(r)}{4r(1-r^3)^2}\left(\omega^2-(1-r^3)k^2\right) = 0\qquad,
\end{equation}
where $\phi =e^{2\pi\imath/3}$ is a cubed root of unity.

This equation is seen to have four singular points, $r_0$, with the following respective characteristic exponents, $\lambda$:
\[
\begin{array}{c||c|c|c|c}r_0&0&1&\phi &\phi^2\\ \hline \lambda & 0,2&\pm\imath\omega/6&\pm\imath\omega/(6\phi)&\pm\imath\omega/(6\phi^2)
\end{array}
\]

\subsubsection{Boundary conditions}

In order to solve, or at least numerically find the quasinormal modes for the above perturbation we need to understand where and why to impose boundary conditions. As discussed above, from equation 
$(\ref{eq:wvmetric1})$ for the world-volume metric, we can deduce the existence of a horizon on the world-volume of the stationary solution at the coordinate $r=1$. For densities and currents living on the world-volume (momenta etc.) $r=1$ is a horizon and appropriate boundary conditions must be applied at this point, as discussed in \cite{CST,SSG}. 

Near this world-volume horizon the local behavior of the perturbations is determined by the characteristic exponents at $r=1$: $\lambda= \pm \imath \omega/6$. Their complex nature implies wave behavior in the coordinates $r$ and $\tau$. At the horizon $r=1$ we should have the boundary condition that we keep only those waves that are infalling, i.e. traveling to $r>1$. A careful analysis is required because $\tau$ diverges at $r=1$, as we will now show. 

We are considering local solutions near the world-volume horizon of the form:
\begin{equation}
\label{eq:epwave}
\epsilon(\tau,r,w) = (1-r)^{\pm \imath\omega/6}\delta(r)e^{\imath\left(kw-\omega\tau\right)}\qquad,
\end{equation}
where $\delta(r)$ is regular at the horizon. Near the world-volume horizon, for $r<1$, this becomes:
\[
\epsilon \approx \exp\left(-\imath\omega\left(\mp \log(1-r)/6+\tau\right)\right)\qquad.
\]
Recalling the definition of $\tau$ we see that its near horizon behavior is:
\[
\tau=\gamma^{-2/3}t-\frac{\log(1-r)}{6}+\frac{1-r}{2\left(\gamma^2-1\right)}+\mathcal{O}\left((1-r)^2\right)\qquad.
\]
This gives the near horizon behavior of the solution:
\[\epsilon\approx\exp\left(-\imath\omega\left(-\frac{\log(1-r)}{6}\left(1\pm1\right)+\frac{1-r}{2\left(\gamma^2-1\right)}+\gamma^{-2/3}t\right)\right)\qquad.
\]
We see that, for the top sign, the logarithmic term is dominant, and following a constant phase, as $t$ increases we require $\log(1-r)$ to increase, that is to move towards smaller $r$, away from the horizon at $r=1$. This means the top sign corresponds to outgoing boundary conditions at the world-volume horizon. For the lower sign, the logarithmic term disappears and as $t$ increases a constant phase requires $(1-r)$ to decrease, that is for $r$ to get closer to the horizon at $r=1$. Therefore the bottom sign corresponds to infalling boundary conditions at the world-volume horizon, and on physical grounds we will use the characteristic exponent $-\imath\omega/6$ as a boundary condition for the perturbations.

Making one last change of dependent function:
\[\epsilon(r)=(1-r)^{-\imath\omega/6}(r-\phi)^{-\imath\omega/(6\phi)}(r-\phi^2)^{-\imath\omega/(6\phi^2)}\eta(r)\qquad\]
and defining the variable $x=1-r$, we obtain the equation of motion for $\eta(x)$:
\begin{align}\label{eq:etaeqmo}
\eta(x)(k^2-\omega^2)&+\eta'(x)4(\imath\omega-3+2(3-\imath\omega)x+(\imath\omega-6)x^2+2x^3)\nonumber \\ &+\eta''(x)4(-3x+6x^2-4x^3+x^4)=0\qquad.
\end{align}

The above discussion of boundary conditions required $\eta$ to be regular at the horizon, so expanding in a local series solution about the horizon at $x=0$, $\eta(x) \equiv \sum_{n=0}^\infty c_n x^n$, we obtain the recursion relation for the coefficients $c_n$, with $n\ge3$:
\begin{equation}\label{eq:recur}
c_n = -\sum_{i=1}^3\frac{c_{n-i}\left(r_0\delta^i_1+(n-i)\left(s_i+(n-i-1)t_{i+1}\right)\right)}{n\left(s_0+(n-1)t_1\right)}\qquad,
\end{equation}
where $r_i$, $s_i$, and $t_i$ denote the coefficients of $x^i$ in equation (\ref{eq:etaeqmo}) of $\eta$, $\eta'$, and $\eta''$, respectively. That is $r_0=k^2-\omega^2$, $s_1=8(3-\imath\omega)$, $t_4=4$, etc. The first three terms of the recursion relation are given by the initial conditions:
\[
c_0 \equiv 1,\quad c_1 = \frac{\omega^2-k^2}{4(\imath\omega-3)}, \quad c_2 = \frac{(\omega^2-k^2)(k^2-\omega^2+8(3-\imath\omega))}{32(\imath\omega-3)(\imath\omega-6)}\qquad,
\]
where $c_0$ is simply chosen as the normalization of the solution to our linear ODE.

At the boundary, $x=1$, the desired boundary condition is that $\eta$ vanishes. Demanding this and the ingoing wave condition at the horizon leads to a discrete set of possible $\omega$, the quasinormal frequencies.

\subsubsection{Longitudinal perturbations}\label{sec:longpert}

For a perturbation in the direction of motion $x$, $\epsilon^\mu = \epsilon \delta^\mu_x$, it can be shown that the action for the perturbation, equation (\ref{eq:pertaction}), is simply $\gamma^2$ times the action of the transverse perturbation. This was also the case for the string in AdS$_5$, as seen in \cite{SSG}, and in fact is true for all the ``analytically special'' cases defined in \cite{JK}: those for which a defect of world-volume dimension $n+2$ is in an AdS space of dimension $d+1$ satisfy $d=2(n+2)$. 

Having an identical action up to scaling, for the longitudinal perturbations the equations of motion, and hence quasinormal frequencies, will be the same as for the transverse case. Properties dealing directly with the action, such as energy and momentum currents, will only need to be rescaled by powers of $\gamma$. The existence of the factor of $\gamma^2$ can be understood in terms of Lorentz transformations. Going from the rest frame of the moving defect to the rest frame of the CFT plasma will lead to different transformations for the two point function of the longitudinal and transverse fluctuations. These correlators can be derived from the gravitational action under the duality, and it is seen that the $\gamma^2$ factor is such to reproduce the correct transformation.

That the factor of $\gamma^2$ is the only difference between the two actions is a much less trivial statement. On the field theory side this is equivalent to saying that in the rest frame of the moving defect the two point functions for transverse and longitudinal fluctuations are equal. As pointed out in \cite{CH} for the case of a quark, this means that the fluctuations of the defect are spherically distributed in this frame, despite the effectively moving plasma background. This is only seen to hold for conformal theories, and is expected to be related to the fact that the drag coefficient is velocity independent in CFTs \cite{CH}, although no general understanding from the field theory side exists at this time.

\subsection{Quasinormal frequencies}
Using the above equation of motion and boundary conditions, we can numerically determine the quasinormal frequencies in a variety of ways. The most straight forward is to numerically solve the equation of motion with the infalling boundary conditions at the world-volume horizon, for various values of $k$ and $\omega$, and then determine for which values of $\omega(k)$ are the solutions normalizable at the AdS boundary. This method is known to be limited in accuracy, especially at large $\omega$ and $k$. A similar approach, with similar problems, is to truncate the power series solution for $\eta$ at a finite number of terms, and using the knowledge of the coefficients from the recursion relations (\ref{eq:recur}), we can also calculate for which values of $\omega(k)$ the solutions are normalizable at the AdS boundary. From past experience with the case of the trailing string in AdS$_5$ a more promising approach is to use continued fractions, and their relation to solutions of recursion relations.

\subsubsection{Continued fractions}
An elegant and accurate method for obtaining the numerical values of quasinormal frequencies implements the use of continued fractions, as discussed in \cite{Starinets}. There, a criteria was used for which the local series solution had an increased radius of convergence, due to the existence of a minimal solution to (\ref{eq:recur}). This criteria was provided by Pincherle's Theorem which states that a minimal solution exists if and only if a certain continued fraction converges. By determining for which values of $\omega$ the fraction converges, the quasinormal frequencies are found.

The recursion relations (\ref{eq:recur}), unlike those of \cite{Starinets}, are four term and therefore Pincherle's Theorem, which only relates to three term relations, is inapplicable. Fortunately, due to \cite{Parusnikov}, a generalization to $m+2$-term recursion relations exist, relating the existence of minimal solutions to the convergence of $m$-dimensional continued fractions. The pertinent results are discussed in the Appendix.

In general, the continued fraction technique takes advantage of the fact that the characteristic exponents at the boundary differ by an integer, and hence the solution that behaves like $r^0$ contains logarithmic terms. Pincherle's theorem, and its generalization, gives us a criteria assuring that the recursion relation (\ref{eq:recur}) has a minimal solution, which in turn assures that the solution has an increased radius of convergence, and hence is analytic at $r=0$. Therefore, for the values of $\omega(k)$ such that a minimal solution to the recursion relation exists, the corresponding solution to the equation of motion is analytic at the boundary, and hence must behave like $r^2$; that is these $\omega$ correspond to Dirichlet solutions at the boundary with infalling boundary conditions at the horizon and hence are the quasinormal frequencies. 

The first few quasinormal frequencies for the infinitely extending membrane in AdS$_7$ with $k=0$ are listed in the table below. From the differential equation it can analytically be shown that they come in pairs with real parts differing by a sign. The quoted significant figures are those which agree after running the continued fraction algorithm seventy and ninety steps (the $a$ in $f^{(a)}$ in the Appendix); except the highest listed mode, for which the working precision of the numerics could not be made accurate enough after ninety steps. 

\[
\begin{array}{c|c}Re[\omega]&Im[\omega]\\ \hline \pm3.8365833596 & -1.9990040306\\ \pm6.51824614&-3.52723336\\ \pm9.1471822&-5.036024\\ \pm11.76121&-6.5403\\ \pm14.369&-8.043\\ \pm16.98&-9.54
\end{array}
\]

\begin{figure}%
\centering
\includegraphics[width=.9\columnwidth]{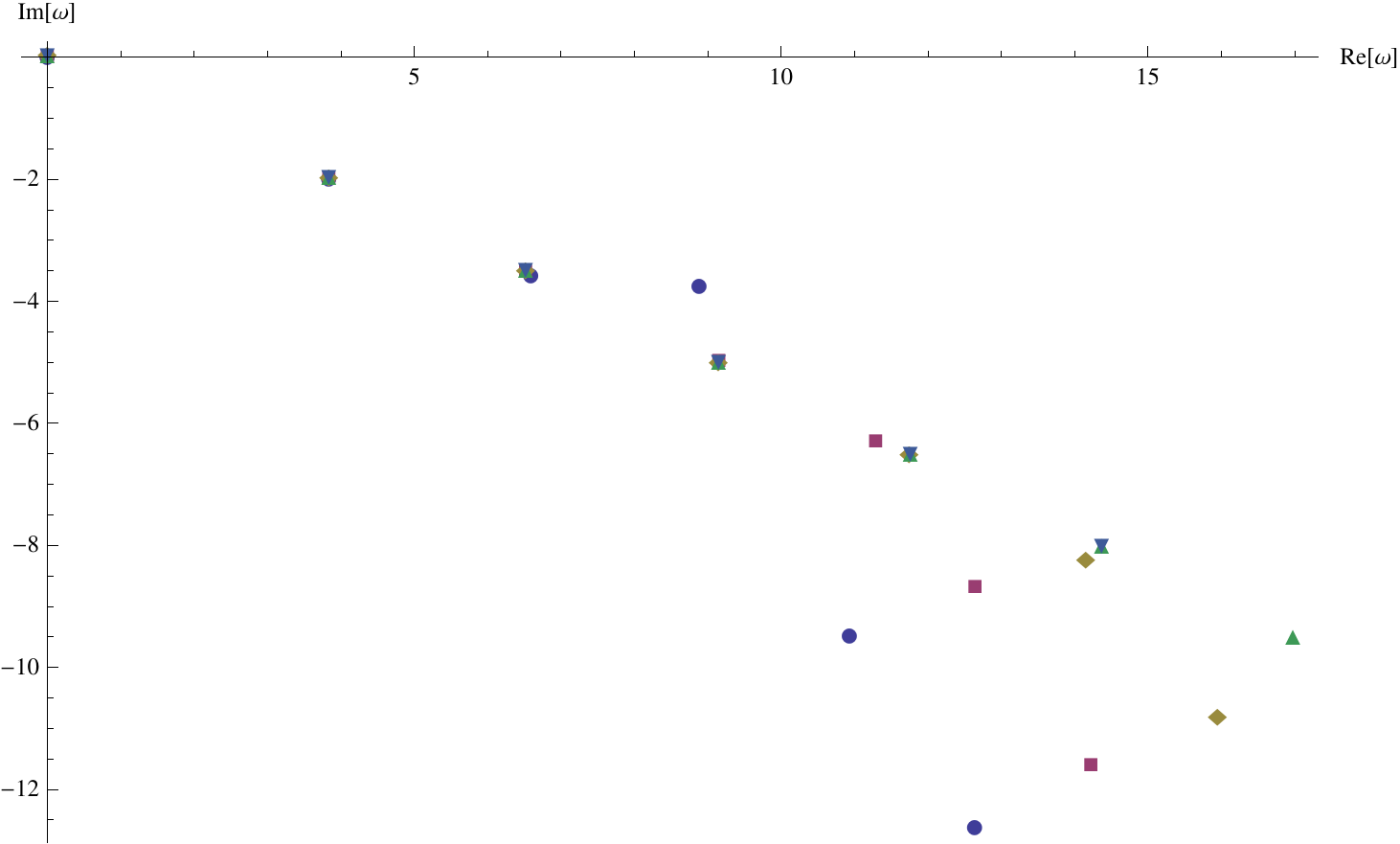}%
\caption{The quasinormal frequencies in the complex $\omega$ plane, as determined numerically by the continued fraction method. Convergence is shown via plotting the frequencies as determined at differing number of steps in the iterative process. Shown are: Circles- 30 steps; Squares- 40 steps; Diamonds- 50 steps; Triangles- 70 steps; Inverted Triangles- 90 steps.}%
\label{fig:qnf}%
\end{figure}

Figure (\ref{fig:qnf}) plots the quasinormal frequencies at $k=0$ for various steps in the numeric iteration, showing the convergence behavior, which for the lower modes is much more accurate than the size of the dots. We note that all frequencies have a negative imaginary part, which indicate a decay of the perturbations and hence stability. The mode at the origin is simply the trivial uniform shift of the stationary solution $x_v$. The structure in the complex plane of a near linear, equally spaced distribution is familiar from the case of perturbations in AdS$_5$ \cite{ANAS}.

\begin{figure}%
\centering
\includegraphics[width=.7\columnwidth]{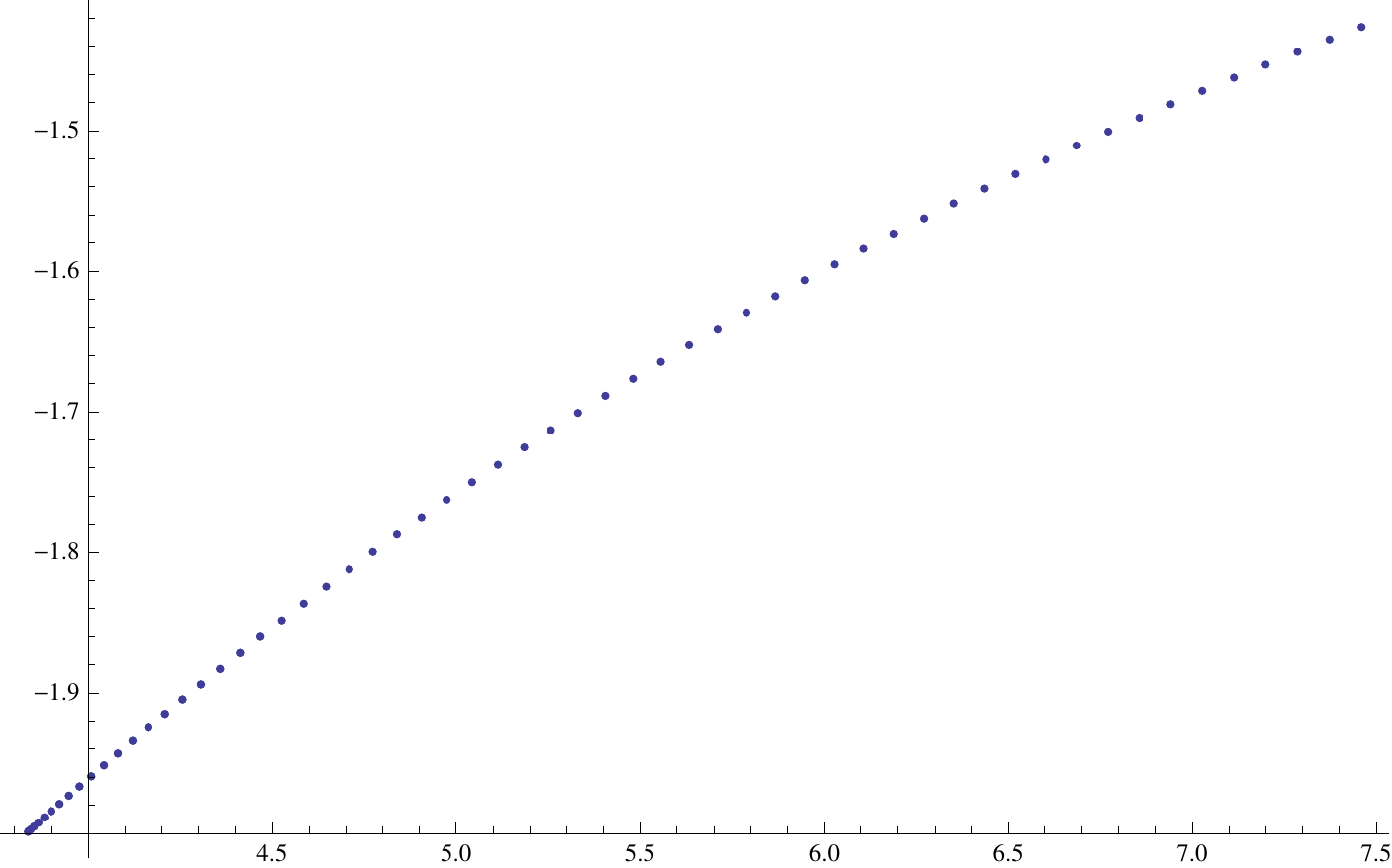}%
\caption{The transverse momentum dependence of the lowest quasinormal mode in the complex $\omega$ plane. As $k$ increases from 0 to 6 the mode moves up and to the right, although appears to stay bounded away from the real axis.}%
\label{fig:kdep}%
\end{figure}

Figure (\ref{fig:kdep}) shows the dependence of the lowest mode on the transverse wave number $k$, for $0\le k\le 6$. This behavior is also qualitatively similar to previous studies.

\section{The Goldstone mode}\label{sec:solu}
\subsection{An analytic solution}
Recalling the discussion above concerning boundary conditions at the world-volume horizon we are led to examine the equation of motion for $\delta(r)$, the regular part of the perturbation that has the infalling wave condition peeled off. Making the ansatz $k=\pm \omega$ for the dispersion relation this equation of motion has the analytic solution:
\begin{align}
\delta(r)=&C_1\exp\left[\frac{\imath\omega}{6}\left(-\sqrt{3}\arctan\left(\frac{1+2r}{\sqrt{3}}\right)+\log\left(\sqrt{1+r+r^2}\right)\right)\right]+\nonumber\\
&C_2\exp\left[\frac{\imath\omega}{6}\left(\sqrt{3}\arctan\left(\frac{1+2r}{\sqrt{3}}\right)+\log\left(\frac{(1-r)^2}{\sqrt{1+r+r^2}}\right)\right)\right]\qquad.
\label{eq:mmode}
\end{align}
We note that this gives an $\epsilon$ that is Neumann at the AdS boundary, $r=0$.
 
\begin{figure}%
\centering
\includegraphics[width=.7\columnwidth]{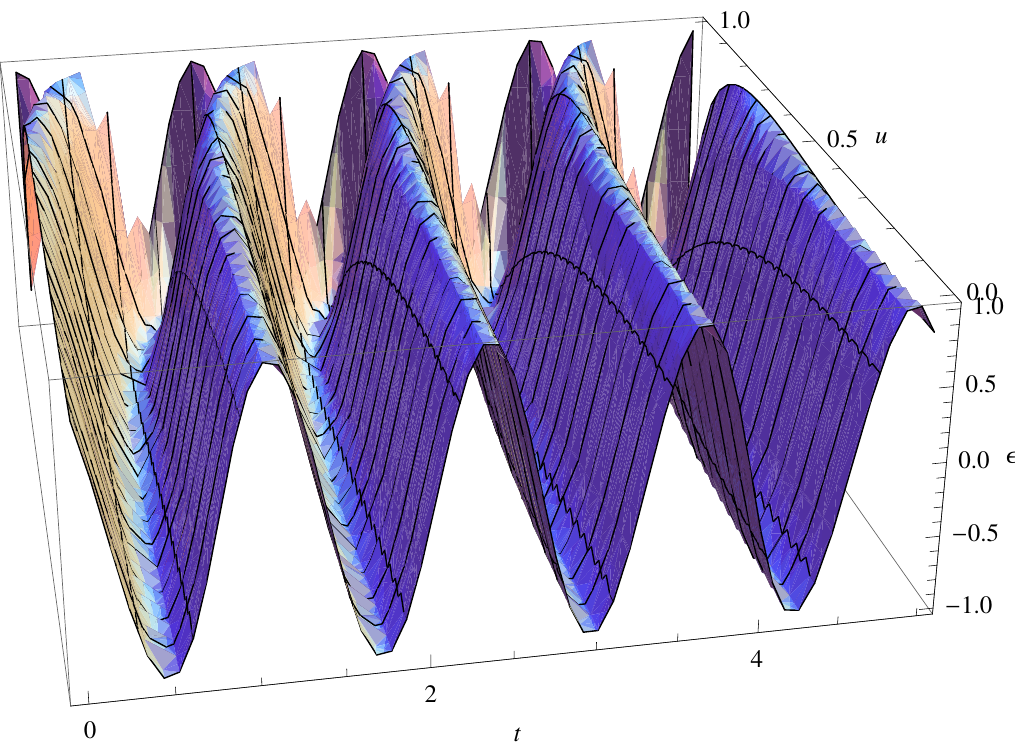}%
\caption{The analytic solution to the perturbation around the uniformly moving membrane. Plotted for $y=0$ and $\omega\gamma^{-2/3}=5$, with $t\in(0,5)$ and $u\in(0,1)$. The boundary is in the foreground at $u=0$; from there the disturbance propagates causally down the membrane and bunches up on the bulk horizon at $u=1$ where the radial wave number becomes infinite.}%
\label{fig:gsmode}%
\end{figure}

Examining the near horizon behavior we see that this solution behaves as:
\[\delta(r\approx 1)=3^\frac{\imath\omega}{12}e^\frac{-\imath\pi\omega}{6\sqrt{3}}C_1+3^\frac{-\imath\omega}{12}e^\frac{\imath\pi\omega}{6\sqrt{3}}C_2(1-r)^\frac{\imath\omega}{3}+\mathcal{O}\left((1-r)^2\right)\qquad.\]
The second term interferes with the desired boundary conditions at the horizon and leads to the unphysical case of outgoing waves, we therefore require $C_2=0$. Normalization then fixes $C_1$. Converting to our original variables, we can write the full perturbation as:
\begin{align}&\epsilon(a)=\nonumber\\
&C_1\cos\left[\omega\gamma^{-2/3}\left(\pm\gamma y-t-\frac{1}{6}\left(\sqrt{3}\arctan\left(\frac{1+2u}{\sqrt{3}}\right)+\log\left(\frac{1-u}{\sqrt{1+u+u^2}}\right)\right)\right)\right]\end{align}
This is plotted in Figure \ref{fig:gsmode} for $y=0$ and $\omega\gamma^{-2/3}=5$. We see that in these field theory coordinates the perturbation has the effective frequency $\omega_{eff}\equiv\gamma^{-2/3}\omega$ and transverse wave number $k_{eff}\equiv\gamma^{1/3}\omega=\gamma\omega_{eff}$. From the boundary the perturbations propagate down the membrane, and bunch up at the bulk horizon $u=1$, this being due to the increasing wave number effect of the logarithmic term $\log(1-u)$.

\subsection{Interpretation}
The dispersion relation in the field theory coordinates is easily interpreted. For a defect at rest in the plasma the dispersion relation is just that for a massless mode, $\omega_{eff}=k_{eff}$. We can understand this as a Goldstone mode: the line defect in the field theory breaks the full translation invariance down to only along the $y$ axis, the direction of its extent. This analytic solution then corresponds to the Goldstone modes for the four broken generators. For the moving defect a similar interpretation in terms of symmetry breaking exists, although the lack of Lorentz invariance due to the preferred rest frame of the plasma makes the dispersion relation less obvious. 

\section{Compact moving membrane}\label{sec:compact}
\subsection{Background solution}
We recall from \cite{JK} the solutions describing a membrane which is a circle on the boundary and moves with a constant velocity in a transverse direction, $x$, perpendicular to the enclosed disk. With this set up we can make the ansatz $x=vt+x(u)$, and $r=r(u)$ for the profile in the direction of motion, and the radius of the defect in the bulk, respectively. Written in traditional Poincar\'e coordinates, $u=z^2$, the equation of motion for $x(z)$ is very similar to that of section \ref{sec:infsolu}, and a constant of first integration allows us to write:
\begin{equation}
x_{,z}^2=\left(\frac{Kvz^3}{1-z^6}\right)^2\frac{\left(\gamma^{-2}-z^6\right)\left(1+(1-z^6)r^2_{,z}\right)}{(1-z^6)r^2-K^2v^2z^6}\qquad.
\label{eq:xcomp}
\end{equation}
Here too the constant $K$ is fixed by demanding $x^2_{,z}$ and $-g$ are everywhere positive on the world-volume. This requires $K=\pm r_c\gamma$, where $r_c\equiv r(z_c=\gamma^{-1/3})$ is the radius of the defect at the critical bulk coordinate, $z_c$, where there is seen to be a world-volume horizon, as above. The sign is determined to be the bottom one, on the physical grounds that energy should be lost into the black hole, and not emerge.

We can now substitute this into the equation of motion for $r$ and obtain a second order nonlinear differential equation, which we numerically solved in \cite{JK} for different values of the velocity, $v$, and the critical radius, $r_c$. For $r_c=0$, the profile must truncate and cap off smoothly at some bulk radial coordinate $z<z_c$, as seen by examining the energy-momentum flux down the world-volume. This gives the profile termed a ``bowl'' in \cite{JK}. For $r_c>0$ the defect stretches all the way to the bulk horizon, $z=1$; solutions dubbed ``tubes.'' For a range of parameters these tubes have unavoidable conical singularities where $r\to 0$, see the discussion of \cite{JK}.

\subsection{World-volume metric and horizon}
For the reparametrization similar to above:
\[\rho\equiv z\gamma^{1/3},\qquad \tau\equiv\gamma^{-2/3}(t+f(z)),\]
we obtain a diagonal world-volume metric with;
\[g_{\tau\tau}=-\left(\frac{R}{\rho}\right)^2(1-\rho^6),\qquad g_{\rho\rho}=\left(\frac{R}{\rho}\right)^2\frac{r^2(1+(1-\gamma^{-2}\rho^6)r^2_{,z})}{(1-\gamma^{-2}\rho^6)r^2-r_c^2v^2\rho^6}\qquad.\]
The horizon at $\rho=1$ is clear in these coordinates. As discussed above, the existence of a world-volume horizon will be used to apply physically acceptable boundary conditions to perturbations about the above numerical solutions. 

\subsection{Perturbations}
\subsubsection{Equation of motion}
We will consider perturbations about the above solutions for the radial embedding function; perturbations of the form $\epsilon^\mu(a)=\epsilon(a)\delta^\mu_r$. Considering $\epsilon$ and its derivatives small, we can use the expansion of the action to second order, equation (\ref{eq:pertaction}), to obtain the equation of motion outside the world-volume horizon:
\begin{equation}
\epsilon''(\rho)+\frac{p'}{p}\epsilon'(\rho)+\left(-m^2\frac{r}{p}-\omega^2\frac{q}{p}\right)\epsilon(\rho)=0\qquad,
\label{eq:compperteqmo}
\end{equation}
where we have used the parametrization and ansatz $\epsilon(a)=\exp[\imath m \phi-\imath\omega \tau]\epsilon(\rho)$, and:
\begin{align}p=&\frac{r^2+r_c^2v^2\rho^6r^2_{,z}}{\rho^3r^2}\sqrt{\frac{(1-\rho^6)((1-\gamma^{-2}\rho^6)r^2-r^2_cv^2\rho^6)}{(1+(1-\gamma^{-2}\rho^6)r^2_{,z})^3}}\qquad,\nonumber\\
\frac{r}{p}=&\frac{\gamma^{-2/3}(1+(1-\gamma^{-2}\rho^6)r^2_{,z})}{(1-\gamma^{-2}\rho^6)r^2-r^2_cv^2\rho^6}\qquad,\nonumber\\
\frac{q}{p}=&-\frac{r^2(1+(1-\gamma^{-2}\rho^6)r^2_{,z})}{(1-\rho^6)((1-\gamma^{-2}\rho^6)r^2-r^2_cv^2\rho^6)}\qquad.
\end{align}

\subsubsection{Boundary conditions and quasinormal modes}
As discussed above for the case of the infinitely extended membrane, the existence of a world-volume horizon gives us a location and rationale for the imposition of boundary conditions. At this horizon we should have infalling boundary conditions on our perturbations. We can understand how this will manifest by examining the equation of motion (\ref{eq:compperteqmo}) near this singular point. A power series for $r$ near $\rho=1$ was determined in \cite{JK} in order to implement the numerics finding the background solutions. This analytic structure can be used to determine the characteristic exponent near the horizon. For $\epsilon\approx(1-\rho)^\lambda$ it is found that:
\[ \lambda=\pm\frac{\imath\omega\gamma^{2/3}}{\sqrt{6}\sqrt{3\gamma^{4/3}+\sqrt{\frac{\gamma^2-1}{r_c^2}+9\gamma^{8/3}}}}\equiv\pm\imath\omega B\]
The complex nature of the exponent leads to wave behavior in the $\rho-\tau$ coordinates, and demanding infalling boundary conditions picks the lower sign.

\begin{figure}%
\centering
\includegraphics[width=.7\columnwidth]{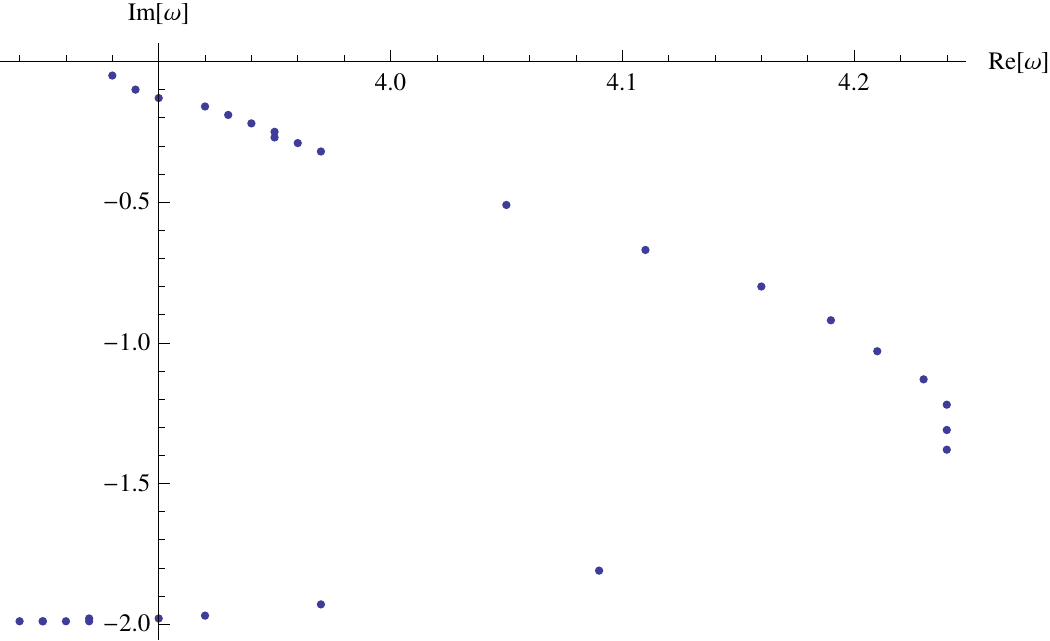}%
\caption{The lowest quasinormal frequency of the tube solutions, with $\gamma=6$ and $m=0$. From top to bottom $r_c$ increases from .002 to 1.1, approximately in logarithmically spaced steps.}%
\label{fig:qnftube}%
\end{figure}

\begin{figure}%
\centering
\includegraphics[width=.7\columnwidth]{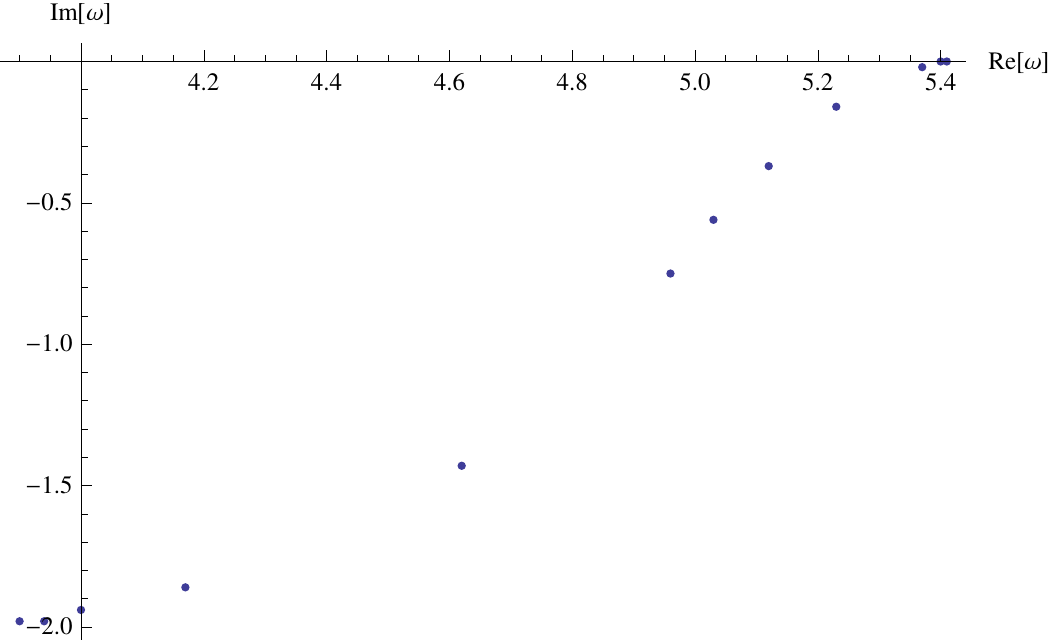}%
\caption{The lowest quasinormal frequency of the tube solutions, with $\gamma=6$ and $m=1$. Again $r_c$ increases from top to bottom.}%
\label{fig:qnfmdep}%
\end{figure}

We now write $\epsilon(\rho)\equiv(1-\rho)^{-\imath\omega B}\eta(\rho)$ where $\eta$ is regular at the horizon. The known power series approximation for $r$ now allows the determination of the power series behavior of $\eta$ near the horizon. As in \cite{JK}, this gives us enough boundary conditions at the horizon to numerically integrate the equation of motion. To determine the quasinormal frequencies we examine the boundary behavior of $\eta$ and numerically determine for which frequencies $\omega(m)$ the solutions have the desired vanishing Dirichlet values. Unlike in section \ref{sec:infsolu}, the numerical nature of our background solution precludes the use of any continued fraction method. The lowest lying quasinormal frequencies are shown in Figure \ref{fig:qnftube} for $\gamma=6$, $m=0$, and various values of $r_c$; Figure \ref{fig:qnfmdep} shows the $m=1$ case. The numerics much beyond these lowest lying modes is severely limited in accuracy. As expected, in the large $r_c$ limit, these frequencies approach that of the infinitely extended sheet of Section \ref{sec:infsolu}. 

Of consequence is the non-positive nature of the imaginary part of the quasinormal frequencies. This indicates that the perturbations decay in time, and that the background tube solutions are classically stable. This comes as somewhat of a surprise, as the situation is reminiscent of the Gregory-Laflamme \cite{GL} situation, where a classical instability is seen for extended objects (in their case black strings) which have a proper radius smaller than the radius of curvature of the AdS space. In our coordinates this corresponds to the radius of the tube of order the bulk radial coordinate $z$; in terms of the radius at the world-volume horizon this translates to $r_c=z_c$, or $r_c=6^{-1/3}\approx0.55$ for the case shown in the figures. Nothing special appears to happen in this regime; although the quasinormal frequencies approach the real axis as $r_c$ decreases, indicating an increase in the characteristic damping time, they do not cross the axis and lead to unstable growing modes.

\section{Conclusion}\label{sec:conclu} 
The major purpose of this paper was the determination of the stability of moving membrane configurations in AdS$_7$. For the infinitely extended ``sheet'' of Section \ref{sec:infsolu} we used a continued fraction method to numerically find that the quasinormal frequencies of the membrane all have a negative imaginary part. Similar conclusions were drawn for the compact ``tube'' solutions of Section \ref{sec:compact} using numerical integration. Perturbations to the compact ``bowl'' solutions are trivially stable in this sense: having Dirichlet conditions at the boundary and regularity conditions at the bottom of the bowl, instead of the infalling wave conditions for the tube, they will have fully real normal modes, as was confirmed with numerics. Thus we deduce the stability of all the classical solutions found in \cite{JK}. This was somewhat unexpected for the case of the tubes, where a Gregory-Laflamme instability was expected. Additionally, the hope of observing hints of turbulent behavior of the plasma flow through an instability of the probes is not borne out.

Two unexpected finds surfaced during this work. The first is the determination of the massless Goldstone modes of Section \ref{sec:solu}. These can be expected to exist on symmetry breaking grounds, but to have found an analytic solution for their behavior is a lucky event. A fuller understanding of their properties would be desirable. The other interesting find was the relation between the actions for the longitudinal and transverse perturbations, discussed in \ref{sec:longpert}. Thus for the analytically special cases of $d=2(n+2)$, we have shown that in its rest frame an $n$-spatial dimensional defect in the $d$-dimensional CFT has a symmetric distribution of its perturbations, despite the presence of an effectively moving plasma.

The last interesting development was the use of higher dimensional continued fractions to calculate quasinormal modes. This approach should be applicable to any set of $k$-term recursion relations, and not just the three-term case that has been most exploited in the literature via Pincherle's Theorem. We believe that this is the first time Parusnikov's generalization has been used in this context, and expect it to have much a greater application.

\section*{Acknowledgments}
I would like to thank Andreas Karch for much discussion and guidance throughout this work. Dam Son and Carlos Hoyos-Badajoz also contributed critically to my understanding. This work was supported in part by the U.S. Department of Energy 
under Grant No. DE-FG02-96ER40956.
 
\appendix
\section{A generalization of Pincherle's theorem}\label{app:contfrac}
In \cite{Parusnikov} the author generalizes Pincherle's theorem to relate $m+2$-term recursion relations and $m$-dimensional continued fractions; that is vectorial continued fractions. This necessitates the use of projective vectors to accomplish ``division by a vector.'' The needed results, along with required definitions, are discussed in the following.

Parusnikov's Theorem 2 assures us the following correspondence:
\[
\mathbf{Theorem \quad2-}
\]
\[
\textrm{iii) $(m+2)$-term recursion relation:}\quad q_{n} = \sum^{m+1}_{j=1} p_{m+2-j,n}q_{n-j}\qquad,
\]
\[
\iff
\]
\[
\textrm{i) interruption-free $m$-dimensional continued fraction:}
\]
\begin{equation}\label{eq:thm2}
\vec{f}_0 =\vec{b}_0+\frac{a_0}{\vec{b}_1+\frac{a_1}{\vec{b}_2+\cdots}}\qquad,
\end{equation}
where $\vec{p}_n\equiv \left(a_n^{-1},a_n^{-1}b_{1,n},\cdots,a_n^{-1}b_{m,n}\right)$; ``interruption-free'' means the $n$-th remainder, $a_n/(\vec{b}_{n+1}+\cdots)$, is non-zero; and throughout the notation $v_{i,j}$ will denote the $i$-th component of vector $\vec{v}_j$.

The construction of higher dimensional continued fractions involves projective vectors, and linear algebra. Given the vector $\vec{p}_n$ defined above, a simple iterative algorithm for numerically calculating continued fraction $\vec{f}_0$ follows:
Define:
\[
\mathbf{P}_n \equiv \left(\begin{array}{c c c c}0&\ldots & 0 &p_{1,n}\\
1&\ldots&0&p_{2,n}\\
\vdots&\ddots&\vdots&\vdots\\
0&\ldots&1&p_{m+1,n}\end{array}\right),\quad
\mathbf{Q}_n\equiv \mathbf{P}_0 \cdots \mathbf{P}_n,\]
\[\mathbf{J}^{-1}\equiv \left(\begin{array}{c c c c c}0&1&0&\ldots&0\\
0&0&1&\ldots&0\\
\vdots&\vdots&\vdots&\ddots&\vdots\\
0&0&0&\ldots&1\\
1&0&0&\ldots&0\end{array}\right)\qquad.
\]

These matrices act on the $m+1$-dimensional projective space, denoted $(f_1:\cdots:f_{m+1})$ with $f_i\in \mathbb{C}$, and we will be interested in the subspace of $m+1$-dimensional vectors whose last component does not vanish:
\[
\mathbb{C}\mathbf{P}^{m+1}_{\star}\equiv \{\vec{f} \in \mathbb{C}^{m+1}: f_{m+1} \ne 0\}\qquad.
\]
Lastly we need the map:
\[
\sigma^{-1}:\mathbb{C}\mathbf{P}^{m+1}_{\star} \to \mathbb{C}^m, \quad \sigma^{-1}(f_1:\cdots:f_{m+1})=\left(\frac{f_1}{f_{m+1}},\cdots,\frac{f_m}{f_{m+1}}\right)\qquad.
\]
We then have:
\[
\vec{f}_0=\lim_{i\to\infty} \sigma^{-1}\mathbf{J}^{-1}\mathbf{Q}_i\vec{e}_{m+1}\qquad
\]
where $\vec{e}_{m+1} \equiv (0:\cdots:0:1)$.

This theorem provides a mapping between recursion relations, such as those arising from differential equations (\ref{eq:recur}), and multidimensional continued fractions. The next theorem provides a powerful result concerning the convergence of a continued fraction and the existence of minimal solutions of the recursion relations.
\[
\mathbf{Theorem \quad6-}
\]
\[
\textrm{An $m$-dimensional continued fraction $\vec{f}_0$ converges.}
\]
\[
\iff\\
\]
\[
\textrm{There exists a minimal subspace of the space of sequences satisfying}\]\[
\textrm{the corresponding recursion relations given in (\ref{eq:thm2}).}
\]
Where we recall that a sequence, $h_n$, is minimal if:
\[
\lim_{n\to\infty}\frac{h_n}{g_n} = 0\qquad,
\]
for all other sequences $g_n$.

Theorems 7 and 8 then give a recursion relation satisfied by the minimal solution, allowing an explicit construction of a basis.
\[
\mathbf{Theorem \quad7/8-}\]
\[
\textrm{The minimal basis obeys:}\]\[
k_{n} = \sum^{m-1}_{j=1}\left(\frac{-f_{j,n+1}}{f_{m,n+1}}\right)k_{n-m+j}+\left(\frac{-f_{m+1,n+1}}{f_{m,n+1}}\right)k_{n-m}\qquad,
\]
The notation here is a bit opaque, and the interested reader is referred to the original work \cite{Parusnikov}; the applicable statement to the case at hand follows. Additionally, this formula corrects a typo in the original, in which the roles of $f_{m+1,n+1}$ and $f_{m,n+1}$ are reversed. This is required to be consistent with the line preceding Equation (23) in \cite{Parusnikov}, as well as with the check of part (B) of Pincherle's theorem following Theorem 8. As a final assurance, this correction is needed for the continued fraction results to agree with other numerical methods.

For the case at hand we can summarize these results into a compact formula used to numerically calculate the quasinormal frequencies. Given the recursion relations (\ref{eq:recur}) and the initial conditions $c_l$ for $l=0,1,2$, then for:
\[
f^{(a)}\equiv \mathbf{J}^{-1}\mathbf{Q}_a \vec{e}_{3}\qquad,
\]
Parusnikov's Theorem 7 gives:
\begin{equation}\label{eq:qnf}
c_2 = \lim_{a\to\infty}\left(\left(\frac{-f^{(a)}_{1}}{f^{(a)}_{2}}\right)c_{1}+\left(\frac{-f^{(a)}_{3}}{f^{(a)}_{2}}\right)c_{0}\right)\qquad,
\end{equation}
for the minimal solutions, subject to the above initial conditions. 

We are interested in the minimal solutions to our equation of motion because these solutions are such that their radius of convergence is increased, and hence they are analytic at the singular point of the boundary, $r=0$. Recalling that the characteristic exponents at the boundary are $(0,2)$, we know near boundary solutions will behave as:
\[y_1(r)\approx r^2+\mathcal{O}(r^3)\qquad\]\[
y_2\approx 1+b_1r+h\log(r)y_1(r)+\mathcal{O}(r^2)\qquad,\]
where $b_1(\omega,k)$ and $h(\omega,k)$ are determined by the recursion relations (\ref{eq:recur}), and the crucial logarithmic term arises due to the characteristic exponents differing by an integer. Thus we see that the solution with non-Dirichlet boundary conditions at $r=0$ is also the solution with a non-analytic logarithmic term. Therefore, the minimal solution, being analytic at the boundary, is also Dirichlet, and given its infalling behavior at the horizon corresponds to a quasinormal mode. This generalization of Pincherle's theorem gives us a criteria for determining those $\omega$ such that a minimal solution to (\ref{eq:recur}) exists, i.e. quasinormal frequencies are those for which (\ref{eq:qnf}) holds. Numerically, we calculate $f^{(a)}$ for large $a$ and then use (\ref{eq:qnf}) to calculate $\omega$. Repeating for larger $a$ gives us an understanding of the convergence properties and allows the approximation of the quasinormal frequencies to a desired accuracy.

\bibliography{bib}{}
\bibliographystyle{elsart-num}

\end{document}